\documentclass{iau} 
\usepackage{graphicx}

\newcommand{\beq}{\begin{equation}}
\newcommand{\eeq}{\end{equation}}
\newcommand{\beqa}{\begin{eqnarray}}
\newcommand{\eeqa}{\end{eqnarray}}
\newcommand{\mnras}{$MNRAS$}
\newcommand{\apj}{$ApJ$}
\newcommand{\aap}{$A\&A$}

\newcommand{\aapr}{$A\&AR$}

\title[Getting started] 
{Getting started: How a supersonic stellar wind is initiated from a hydrostatic surface}

\author[Stan Owocki]   
{Stan Owocki
}

\affiliation{Department of Physics \& Astronomy, Bartol Research Institute, 
University of Delaware, Newark, DE 19716 USA \\ email: {\tt owocki@udel.edu}}

\pubyear{2022}
\volume{366}  
\jname{The Origin of Outflows from Evolved Stars}
\editors{Leen Decin\& Clio Gielen, eds.}
\begin{document}

\maketitle

\begin{abstract}
Most of a star's mass is bound in a hydrostatic equilibrium in which pressure balances gravity. But if at some near-surface layer additional outward forces overcome gravity, this can transition to a supersonic, outflowing wind, with the sonic point, where the outward force cancels gravity, marking the division between hydrostatic atmosphere and wind outflow. This talk will review general issues with such transonic initiation of a stellar wind outflow, and how this helps set the wind mass loss rate.
The main discussion contrasts the flow initiation in four prominent classes of steady-state winds: 
(1) the pressure-driven coronal wind of the sun and other cool stars;
(2) line-driven winds from OB stars;
(3) a two-stage initiation model for the much denser winds from Wolf-Rayet (WR) stars;
and (4) the slow ``overflow'' mass loss from highly evolved giant stars.
A follow on discussion briefly reviews  eruptive mass loss, with particular focus on the giant eruption of $\eta$\,Carinae.
\end{abstract}

\firstsection 
\section{Introduction}

\begin{figure}
\begin{center}
\includegraphics[width=4.in]{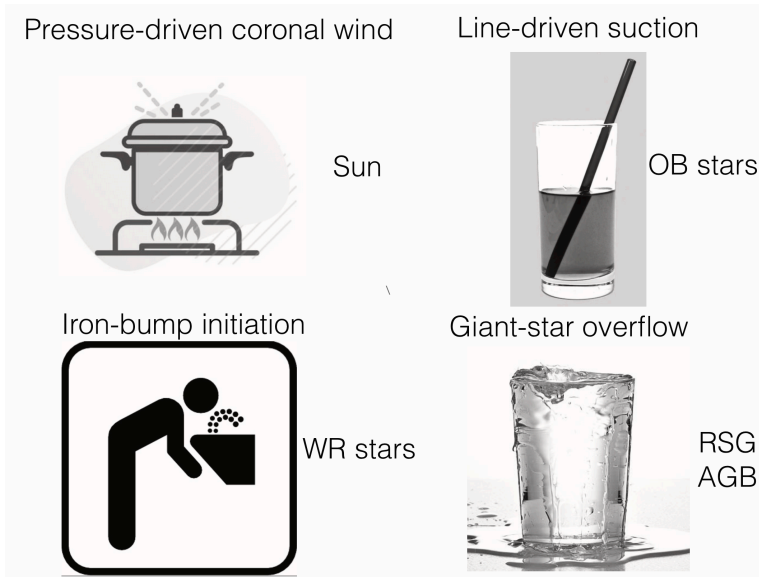} 
 \caption{Icons to represent analogies for processes initiating the four different kinds of steady stellar wind outflow.}
\label{fig01}
\end{center}
\end{figure}

\begin{figure}
\begin{center}
\includegraphics[width=5.in]{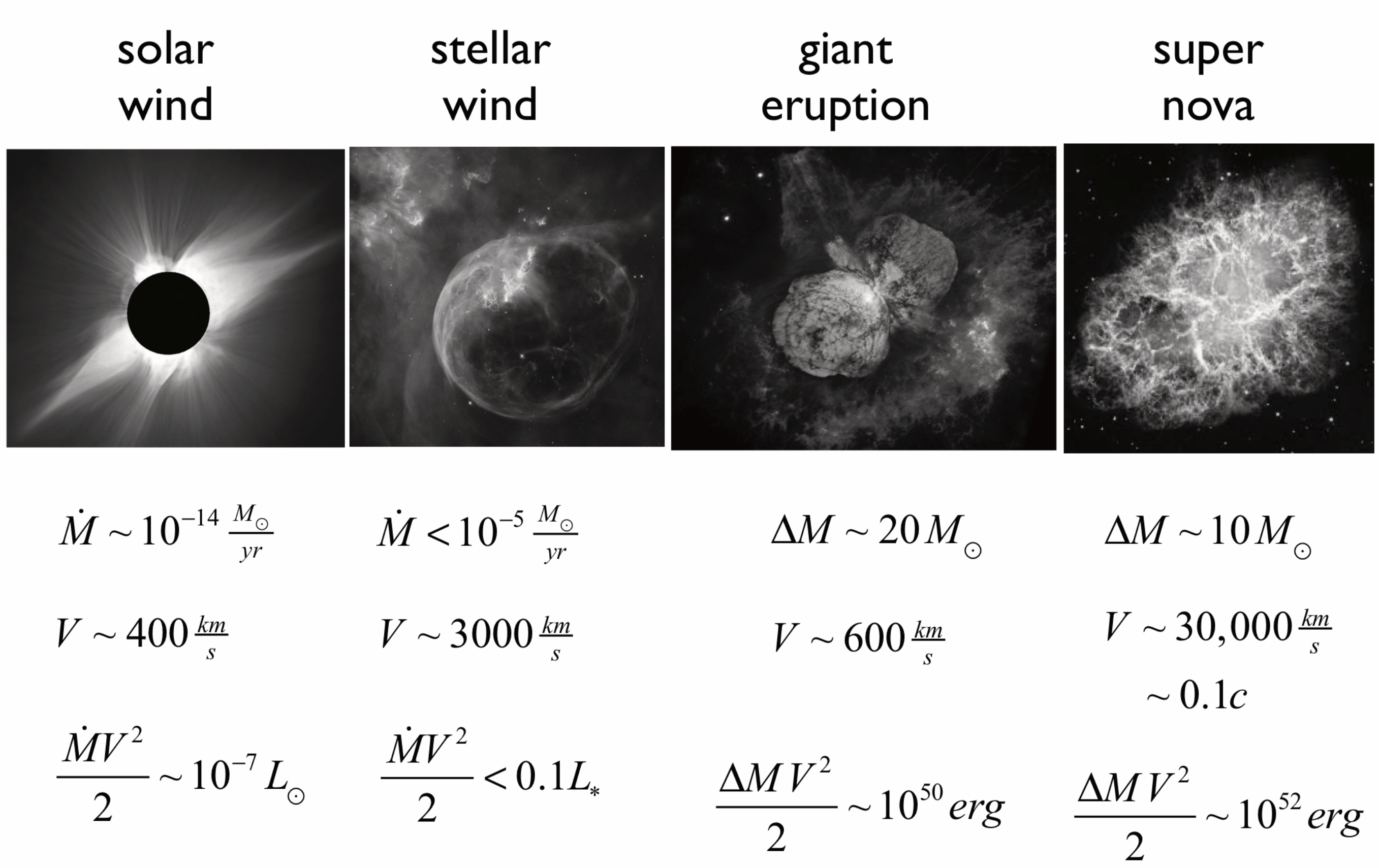} 
 \caption{Comparison of the properties of solar and stellar steady wind outflows (left) with the example of eruptive mass loss from 
 $\eta$\,Carinae (right center), and contrasting that with the explosive mass ejection from a supernova. (right)}
 \label{fig02}
\end{center}
\end{figure}

To set the stage for this symposium's exploration of ``The Origin of Outflows from Evolved Stars", I have been asked to review the basic processes underlying the {\em initiation} of such outflows.
Most of a star's mass is bound in a hydrostatic equilibrium for which the outward push of pressure balances the inward pull of gravity. But if at some near-surface layer additional outward forces overcome gravity, this can transition to a supersonic, outflowing wind, with the sonic point, where the outward force cancels gravity, marking the division between hydrostatic atmosphere and wind outflow. This summary reviews general issues with such transonic initiation of a stellar wind outflow, and how this helps set key wind properties like the mass loss rate and wind flow speed.

As summarized in figure \ref{fig01}, much of the discussion, given in \S 2, focuses on four distinct types of steady-state outflows, namely: 
(1) the pressure-driven coronal wind of the sun and other cool stars;
(2) line-driven winds from OB stars;
(3) a two-stage initiation model for the much denser winds from Wolf-Rayet (WR) stars;
and (4) the slow ``overflow'' mass loss from highly evolved giant stars.
A follow on discussion in \S 3  briefly reviews  eruptive mass loss (see figure \ref{fig02}), with particular focus on a binary merger model for giant eruption of $\eta$\,Carinae, perhaps the most famous of the class of eruptive Luminous Blue Variables (eLBV).

\begin{figure}
\begin{center}
\includegraphics[width=5.0in]{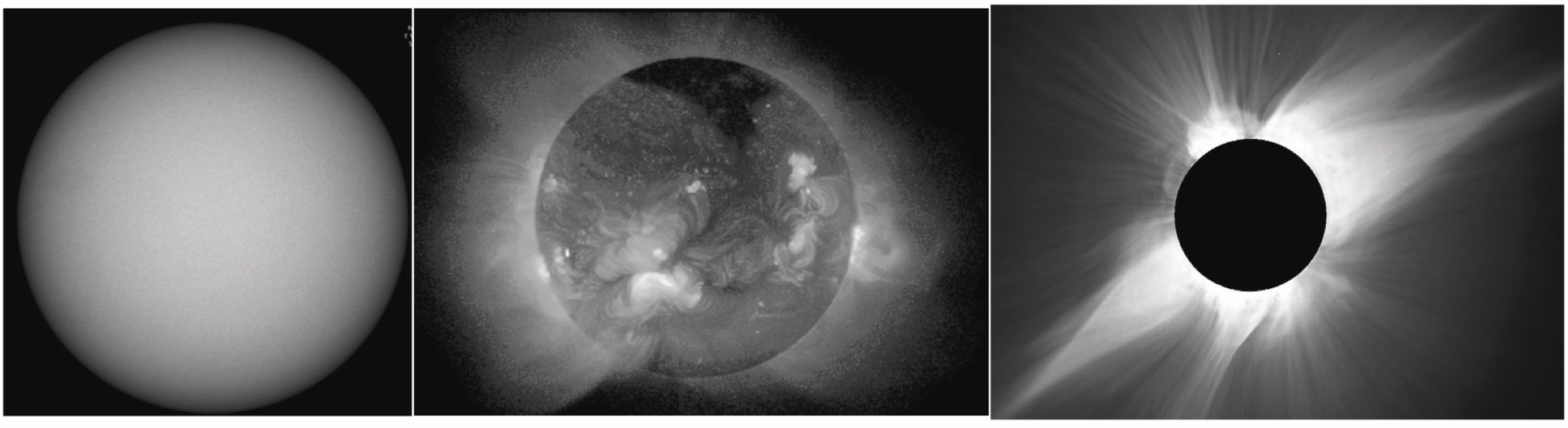} 
 \caption{Images of the solar disk. Left: White light showing solar photosphere; Middle: X-rays showing magnetically confined hot coronal loops ; Right: Solar eclipse showing how coronal loops are stretched outward into pointed ``helmet'' streamers by the expansion into the solar wind}
   \label{fig03}
\end{center}
\end{figure}

\section{Initiation of steady winds}
The mass $M(r)$ within a local radius $r$ of a spherically symmetric star exerts an inward gravitational acceleration $g=G M(r)/r^2$, with $G$ the gravitation constant;
for local mass density $\rho$, this is balanced by the acceleration associated with a pressure gradient $-(1/\rho) dP/dr$.
Using the ideal gas law to write the pressure $P=\rho kT/\mu$ in terms of temperature $T$, molecular weight $\mu$, and Boltzmann's constant $k$, we find the pressure drops exponentially with a characteristic pressure scale height,
\beq
H \equiv \frac{P}{|dP/dr|} = \frac{kT}{\mu g} = \frac{a^2}{v_{orb}^2} R
\, .
\label{eq:Hdef}
\eeq
Here the last equality casts the value at surface radius $r=R$ in terms of ratio of the (isothermal) sound speed $a \equiv \sqrt{kT/\mu}$ to surface orbital speed $v_{orb} \equiv \sqrt{GM/R}$.
For the solar photosphere, we find $a \approx 10$\,km/s while $v_{orb} \approx 420$\,km/s, thus implying a scale height 
that is a tiny fraction $H/R \approx 4 \times 10^{-4}$ of the solar radius.
This is the essential reason the edge of the solar disk appear so sharp in white light images, such as shown in the leftmost panel in figure \ref{fig03}.

Indeed, main sequence stars are all characterized by a similarly small ratio $H/R$, implying that some other force must kick in to overcome gravity and drive a wind outflow.
In the luminous OB and WR stars, this stems from the momentum of scattered radiation, as discussed in \S\S 2.2 and 2.3.

However, for evolved giants, including Red Giants (RG), Red Super-Giants (RSG) and Asymptotic Giant Branch (AGB) stars,
the much weaker surface gravity implies a less tiny ratio for $H/R \gtrsim 0.1$, making it easier for internal variations to induce mass loss,
as discussed in \S 2.4.

\subsection{Pressure-driven coronal winds}
The initial prototype for wind mass loss came from the realization by G. Parker that the high (MK) temperature of the solar corona
would lead to a supersonic expansion we now know as the solar wind.
For a characteristic coronal temperature  of 2\,MK, we find the scale height ratio is now $H/R \approx 0.14$.
The X-ray image in the middle panel of figure \ref{fig03} shows how the hot corona thus has a much greater extension above the solar surface and beyond the solar limb.

If this coronal temperature is kept high -- through extended heating and outward thermal conduction --, the radial drop of gravity implies  this ratio increases outward; 
as such, the pressure no longer continues to drop exponentially, but rather asymptotically approaches a finite value, $P_\infty$, at large radii, $r \rightarrow \infty $.
Relative to the initial pressure $P_o$ at the coronal base, the total drop in pressure for a hydrostatic, isothermal corona is given by
\beq
\frac{P_o}{P_\infty} \approx e^{R/H} ~~ ; ~~ \log \frac{P_o}{P_\infty} \approx \frac{6}{T/{\rm MK}}
\, .
\label{eq:pobpinf}
\eeq
The latter equality shows the pressure drops by 6 decades for $T=1$\,MK, and only 3 decades for $T=2$\,MK.
By comparison, from the solar transition region at the coronal base to the interstellar medium, the 
pressure drop is  actually much greater, $\log (P_{tr}/P_{ism}) \approx 12$.
The upshot is that an extended, hot corona can {\em not} be maintained in hydrostatic equilibrium;
instead, as shown by the outward streamers from the eclipse image in figure \ref{fig03}, it must undergo an outward, supersonic 
{\em expansion} known as the solar wind.

\begin{figure}[b]
\begin{center}
\includegraphics[width=4in]{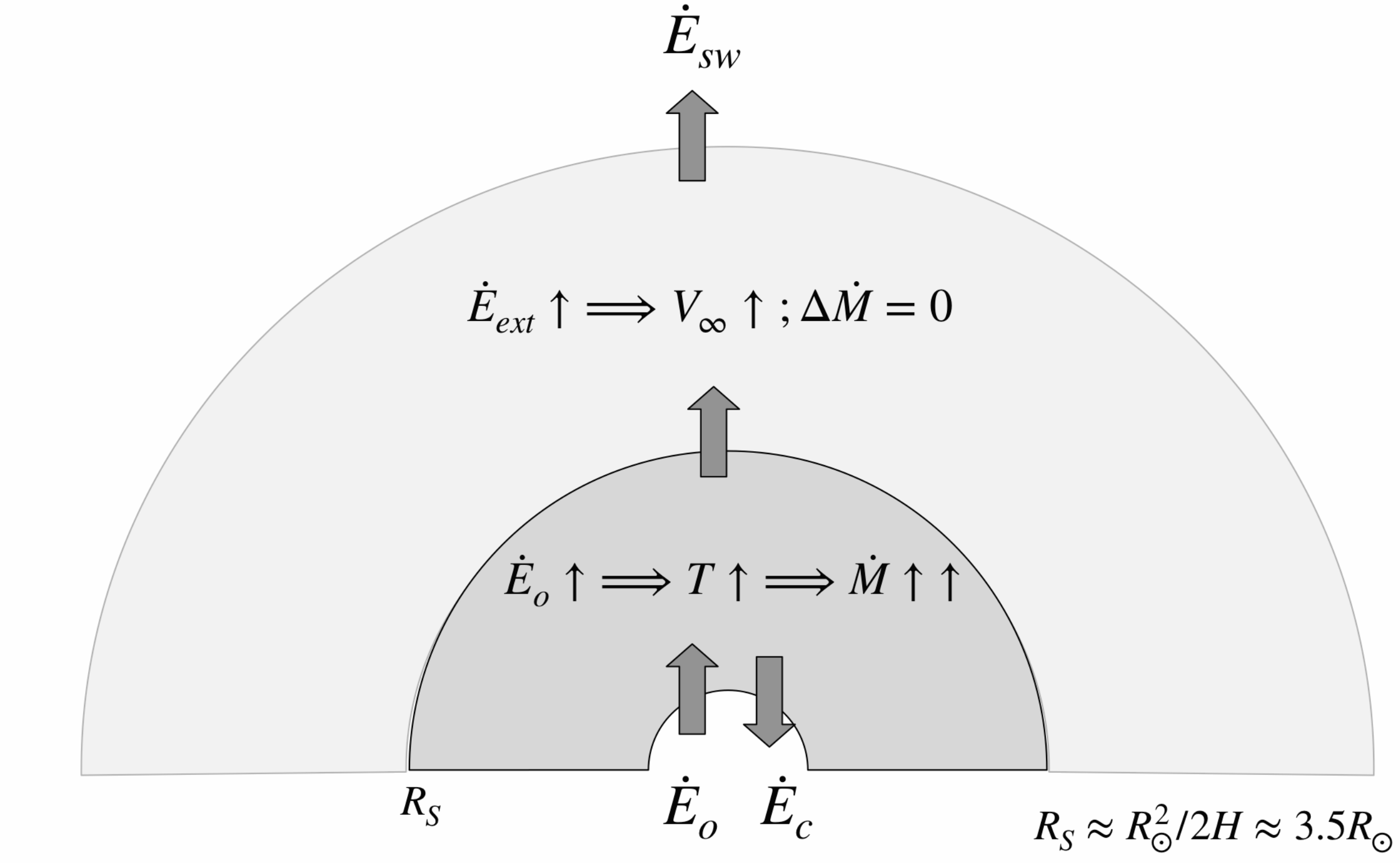} 
 \caption{Schematic to illustrate energy input into the nearly hydrostatic, subsonic coronal base vs. that into the supersonic wind
 above the Parker sonic radius $R_s$.
 For the former, the net input vs. loss by conduction back into the underlying atmosphere leads to coronal heating that sets
 the coronal temperature and location and density at the sonic radius, thus fixing the associated wind mass loss rate.
 For the latter, any further energy addition increases the wind flow speed, with the wind expansion providing the primary mechanism
 to carry out the total net amount of coronal heating.}
   \label{fig04}
\end{center}
\end{figure}

As illustrated by the upper left panel of figure \ref{fig01}, this solar wind expansion can be thought of as analogous to the release valve of a pressure cooker, driven fundamentally by mechanical heating generated by magnetic turbulence in the underlying solar atmosphere.
As illustrated in figure \ref{fig04}, some of this upward energy flux is lost back to the solar atmosphere through thermal conduction,
but the net effect leads to a thermal runaway that raises the coronal temperature to  millions of Kelvin, several hundreds times higher
than the photospheric temperature $T \approx 5800$\,K.

To account for the advective acceleration $v (dv/dr)$ in a spherically expanding wind, one can use mass continuity to split the pressure gradient force into terms that scale with this acceleration and the sphericity, yielding a steady-state equation of motion,
\beq
\left (v - \frac{a^2}{v} \right ) \frac{dv}{dr} = 
-\frac{GM}{r^2} + \frac{2 a^2}{r}
\, .
\label{eq:solwineom}
\eeq
In the subsonic region $v \ll a$, this reduces to the condition for hydrostatic equiliibrium;
but at a critical (``Parker") radius, 
\beq
R_s 
= \frac{GM_\odot}{2a^2} 
= \frac{R_\odot^2}{2H} 
= 3.5 R_\odot  \left ( \frac{2{\rm MK}}{T} \right )
\, ,
\label{eq:Rs}
\eeq
where the RHS vanishes, 
the hydrostatic coronal base transitions  to a supersonic outflow, driven by  the high gas pressure associated 
with the high coronal temperature.

The density and radius of this sonic point set the wind mass loss rate ${\dot M} = 4 \pi \rho_s a R_s^2$, with typical values ${\dot M} \approx 10^{-14} M_\odot$/yr.
Physically, this is set by the level and location of the coronal heating.
Heat added within the subsonic coronal base increases the scale height, with less drop in density toward a closer sonic point, and thus 
a direct increase in ${\dot M}$.
In contrast, extended heating into the supersonic wind, where the mass loss rate is already fixed, instead leads to higher energy per unit mass, 
and thus a higher wind speed.

While the detailed mechanisms for coronal heating remain a subject of much current research,
the thermal runaway and associated coronal expansion are thought to be quite robust consequences of the turbulence generated in all cool
stars ($T_\ast < 10,000$\,K) with convective envelopes associated with the opacity blockage from Hydrogen recombination.
The low coronal density needed to avoid strong radiative cooling limits the mass loss rates to values, of order $10^{-14} M_\odot$/yr, that do not appreciably reduce the stellar mass over evolutionary timescales. But the enhanced angular momentum loss associated with a global magnetic field can lead to an effective spindown of the stellar rotation.

\subsection{Line-driven winds from OB stars}
In more massive, hotter OB stars with surface temperatures $T_\ast > 10,000$\,K hydrogen remains ionized up to the surface; such hot stars thus lack the H-recombination convection zones and associated the magnetic turbulence that heats the hot corona,  and associated pressure-driven expansion of cool star winds.
However such hot stars have a much higher radiative luminosity, and the outward force from line scattering of this radiation can overcome gravity and so drive a {\em line-driven} stellar wind outflow.
For opacity $\kappa_\nu$ at a frequency $\nu$ with radiative flux $F_\nu$, the total radiative acceleration depends on the frequency integral,
\beq
g_{rad} = \int_0^\infty d\nu \, \frac{\kappa_\nu F_\nu}{c} = \frac{\kappa_e F}{c}
\,
\label{eq:grad}
\eeq
where the last equality applies in the simplest case of continuum scattering by free electrons,
with $F$ the bolometric flux.

For  a fully ionized gas with solar  hydrogen mass fraction $X=0.72$,
such electron scattering has an opacity $\kappa_e = 0.34  \, {\rm cm}^2$/g.
The ratio of the associated radiative acceleration to gravity is given by the Eddington parameter,
\beq
\Gamma_e \equiv \frac{\kappa_e F/c}{g} = \frac{\kappa_e L}{4 \pi GMc}  \approx \frac{M}{200 M_\odot} 
\, ,
\label{eq:game}
\eeq
wherein the inverse-square radial dependence of both the radiative flux $F=L/4\pi r^2$ and gravity $g=GM/r^2$ cancels, showing this Eddington parameter depends only  on the ratio $L/M$ of luminosity to mass.
The last equality applies for the standard radiative envelope scaling $L \sim M^3$, and provides a rationale for the upper limit to stellar mass, 
which is empirically found to be around $200 M_\odot$.
Stars that approach or exceed the Eddington limit $\Gamma_e = 1$ can have strong eruptive mass loss, as discussed in \S 3 for  
eruptive Luminous Blue Variable stars like $\eta$\,Carinae.

 \begin{figure}[t]
\begin{center}
\includegraphics[width=5.5in]{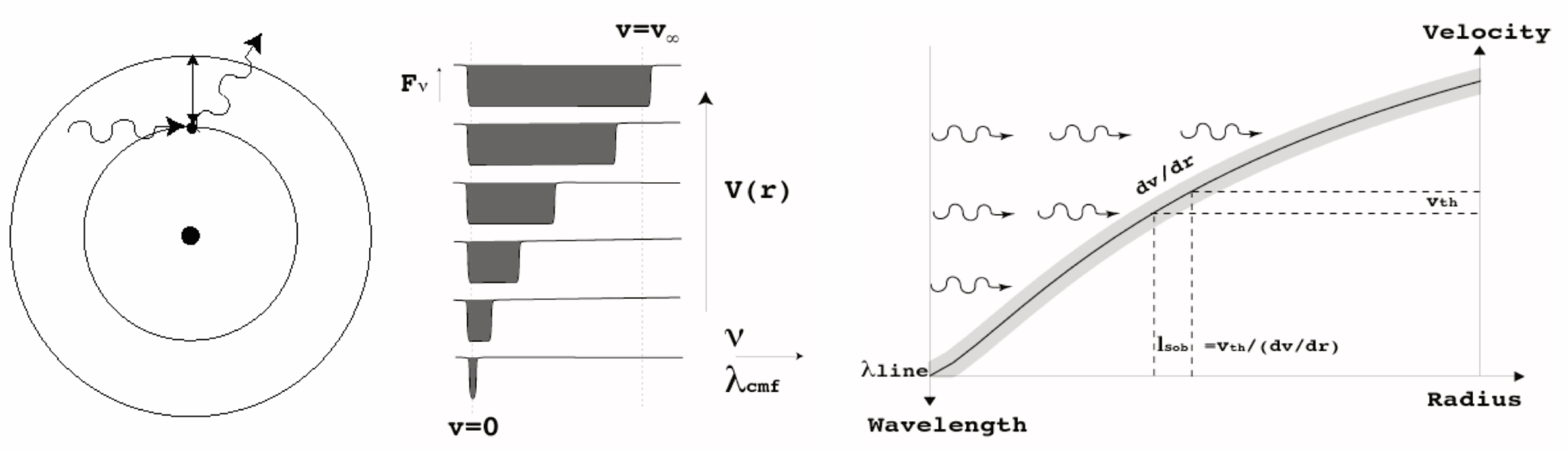} 
 \caption{Left: Illustration of resonant scattering between two bound levels of an ion.
 Middle: The doppler-shifted desaturation that spreads line absorption over a large range of frequencies, ranging up to the frequency associated with the doppler shift from the wind terminal speed, $v_\infty$.
 Right: Continuum photons propagating to the doppler-shifted line resonance that is thermally broadened by over a Sobolev length $l_{sob} = v_{th}/(dv/dr)$, where $v_{th}$ is the ion thermal speed.
 }
   \label{fig05}
\end{center}
\end{figure}

But the steady winds from luminous stars with $\Gamma_e \gtrsim 10^{-3}$ are understood to be driven by the {\em line} scattering from electrons bound into heavy ions ranging from CNO to Fe and Ni.
As illustrated in the leftmost panel of figure \ref{fig05}, the resonance nature of such bound-bound line-scattering greatly enhances the opacity, by a factor ${\bar Q} \gtrsim 10^3$ (\cite[Gayley 1995]{Gayley95}), thus making it possible to overcome gravity and drive a wind outflow for stars with 
$\Gamma_e \gtrsim 1/{\bar Q} \approx 10^{-3} $.

In practice this maximal line acceleration from {\em optically thin} scattering is reduced by the saturation of the reduced flux within an optically thick line.
But as  illustrated in the middle panel of figure \ref{fig05},  the doppler shift associated with the wind acceleration acts to {\em desaturate}
this line absorption, effectively sweeping the absorption through a broad frequency band extending out to the frequency associated with the doppler shift from the wind terminal speed, $v_\infty$.

The right panel of figure \ref{fig05} illustrates how the wind doppler shift of line resonance concentrates the interaction of continuum photons
into a narrow resonance layer with width set by the Sobolev length,  $\ell \equiv v_{th}/(dv/dr)$ (\cite[Sobolev(1960)]{Sobolev60}), associated with 
acceleration through the ion thermal speed $v_{th}$ that broadens the line profile.
In the outer wind where $\ell \ll r$, the line acceleration for optically thick lines is reduced by $1/\tau$, where the Sobolev optical depth $\tau \equiv {\bar  Q}  \kappa_e \rho \ell$, giving then a line acceleration $\Gamma_{thick} \sim (1/\rho) (dv/dr)$ that itself scales with the wind acceleration.
Within this Sobolev approximation, \cite[Castor, Abbott and Klein (1975; hereafter CAK)]{CAKI75} developed a formalism that accounts for
the  radiative acceleration from an ensemble of thick and thin lines, deriving thereby scalings for the wind speeds and associated wind mass loss rate.
Much as in the solar wind, the terminal wind speeds scale with the surface escape speed $v_{esc} \equiv \sqrt{2GM/R}$, with values up to $v_\infty \approx 2000$\,km/s.
However, the mass loss rates can range up to ${\dot M} \sim 10^{-5} M_\odot$/yr, and so up to a {\em billion times} that of the solar wind!

\begin{figure}[b]
\begin{center}
\includegraphics[width=5.5in]{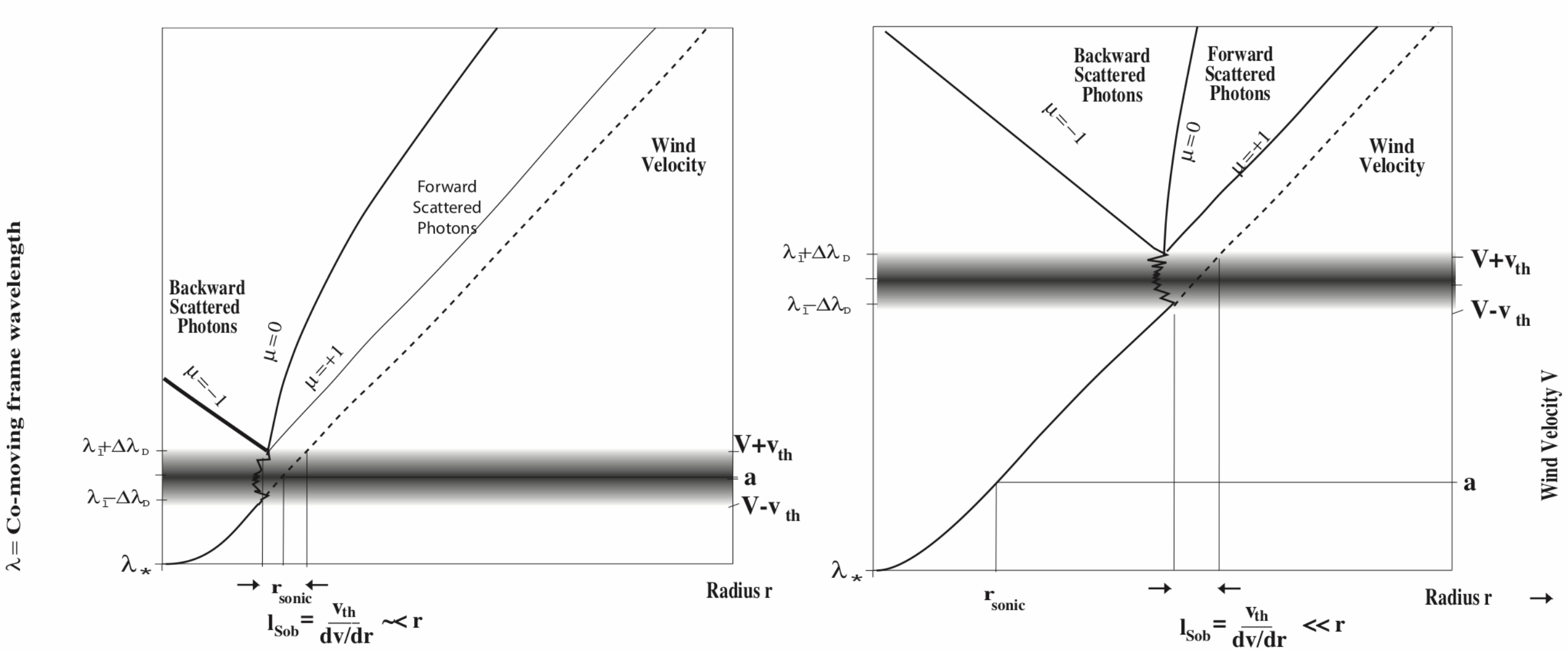} 
 \caption{Alternative rendition of doppler shift from wind acceleration, now depicted as redshift of stellar photons until the come into resonance with a thermally broadened line.
 This allows depiction of the forward and backward scattering of photons escaping this layer.
 In the outer, supersonic wind of the right panel, this is fore-aft symmetric, canceling any net recoil from the diffuse radiation.
 But for inner, transonic wind of the left panel, there can be a net asymmetry between the forward/background scattering, leading then to a non-zero diffuse line force.
 The consequences of this are illustrated in figure \ref{fig07} and discussed in the text.
 }
   \label{fig06}
\end{center}
\end{figure}

A longstanding issue for this Sobolev-based CAK model regards the wind initiation near wind sonic point.
The effective desaturation of a Sobolev model requires,
\beq
\frac{v_{th}}{dv/dr} \equiv  \ell \ll  H 
\approx  \frac{\rho}{|d\rho/dr|}
 \approx \frac{v}{dv/dr}
\, .
\label{eq:ellvsH}
\eeq
At the sonic point $v=a$, this requires $v_{th} \ll a$.
Fortunately, because line driving is by heavy ions with $v_{th}/a \approx 0.3$,  this condition is marginally satisfied;
but it does indicate that a more careful treatment is warranted for the radiative transfer and line driving in this transonic region.

For this it is important to recognize that,  as illustrated in figure \ref{fig06}, line-transfer actually occurs mainly via {\em scattering} not pure absorption.
The right panel shows that, in the highly supersonic outer winds, the escape of photons from the Sobolev resonance is nearly fore-aft symmetric, canceling their recoil, so that the net line-force is the same as if the photon had been purely absorbed.
The left panel shows, however, that in the inner, transonic region this scattering can become asymmetric, with a net recoil and so a non-zero {\em diffuse} line-force.

\begin{figure}[b]
\begin{center}
\includegraphics[width=5.5in]{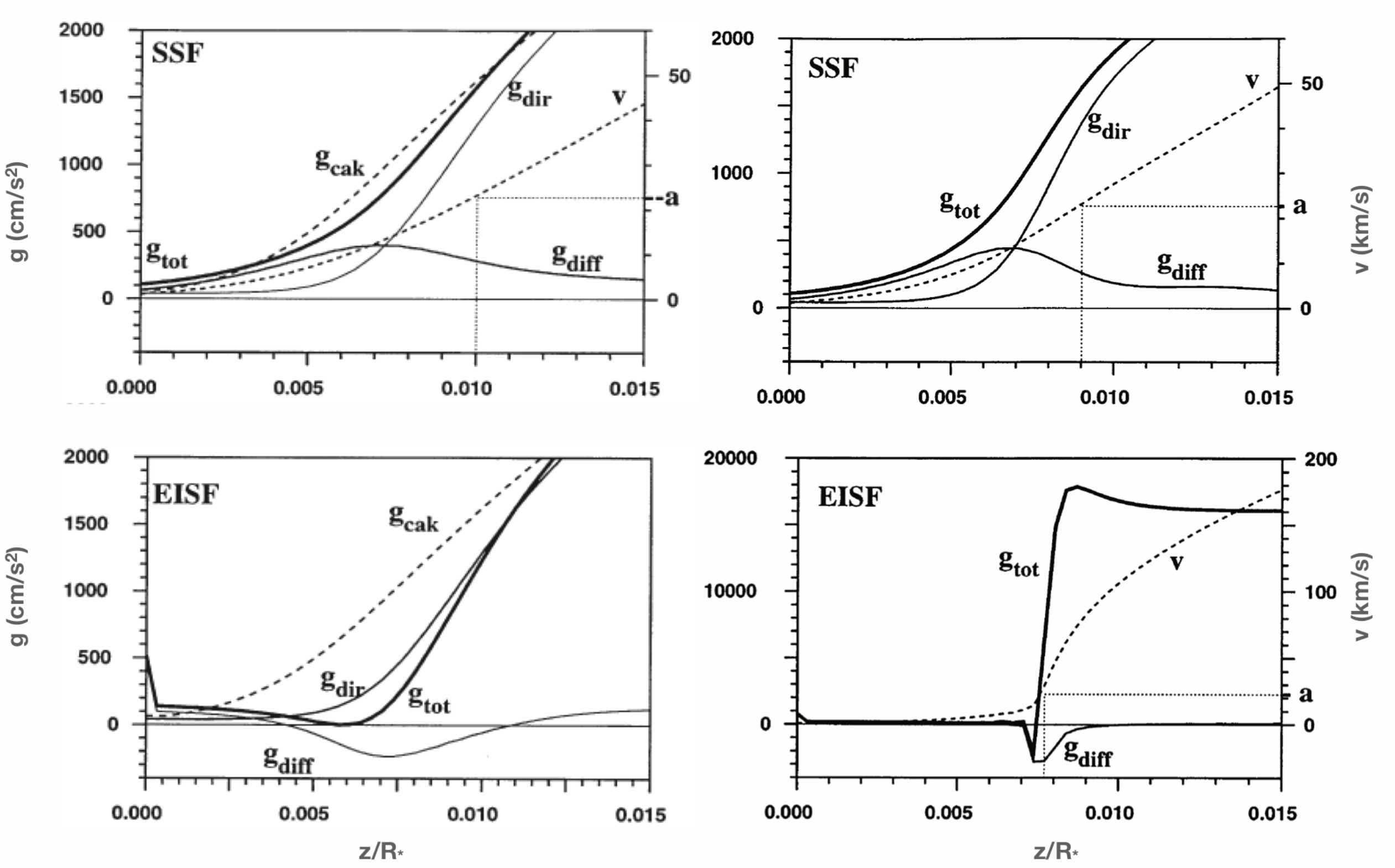} 
 \caption{Radiative acceleration with scattering in the transonic regino (left) and the result initiation of wind velocity,
 for both Smooth Source Function (SSF; top) and Escape Integral  Source Function (EISF; bottom) models of the scattering line transfer.}
   \label{fig07}
\end{center}
\end{figure}

Figure \ref{fig07} compares the radiative driving (left) and resulting wind velocity (right) for two different approximations for the transonic scattering line transfer (\cite[Owocki \& Puls 1996, 1999]{OP96, OP99}).
The top row shows a {\em Smooth Source Functon} (SSF) model, which assumes coupling with the continuum can keep the line source function nearly constant through the sonic region;
the difference in fore/aft escape leads to a {\em positive} diffuse line force, which effectively compensates for the reduction in direct line force associated with the incomplete desaturation in this transonic region. The net result is that the wind driving and velocity are very similar to the standard CAK model.

The bottom row shows an {\em Escape Integral Source Function} (EISF) model, applicable for lower-density cases without sufficient continuum coupling to keep the source function smooth in the transonic region.
The greater escape from the increasing velocity gradient leads now to a marked dip in the source function, with a weaker or even inward diffuse line force that further reduces the net line driving the subsonic region, and thus now a sharp, step-like jump in wind velocity around the sonic point.

An overall point is that in all these models the onset of line-driving near the sonic point represents an effective {\em line-driven suction}, which draws up mass from the underlying hydrostatic equilibrium of the subsonic region.
The reduction in pressure from the outer line-driving induces the underlying subsonic region to expand upward, in much the way that,
 as illustrated in figure \ref{fig01}, the suction on a straw draws up liquid from a glass. This outside-in suction contrasts with the inside-out thermal expansion of a pressure cooker, and of the analogous gas pressure-driven solar wind.

\subsection{Iron-bump initiation of WR winds}
For the much higher mass loss rates of Wolf-Rayet stars, the winds can themselves become optically thick, with the overall wind blanketing increasing the temperature of the transonic region well above the stellar effective temperature, so bringing it closer to the temperature $T \approx 200,000$\,K for a strong opacity bump associated with the huge number of overlapping iron lines.
In OB stars, this ``iron opacity bump" induces a narrow sub-layer of convection that generates gravity waves and the associated surface macro-turbulence inferred from broadening of photospheric spectral lines.
But in WR stars close to the Eddington limit, the inefficiency of the more near-surface convection can instead cause the net force to exceed the Eddington limit, leading to either an inflation of the stellar envelope (\cite[Graefener et al. 2012]{Graefener12}),
or even initiation of an outflow (\cite[Poniatowski et al. 2021]{Poniatowski21}).

\begin{figure}[b]
\begin{center}
\includegraphics[width=5.5in]{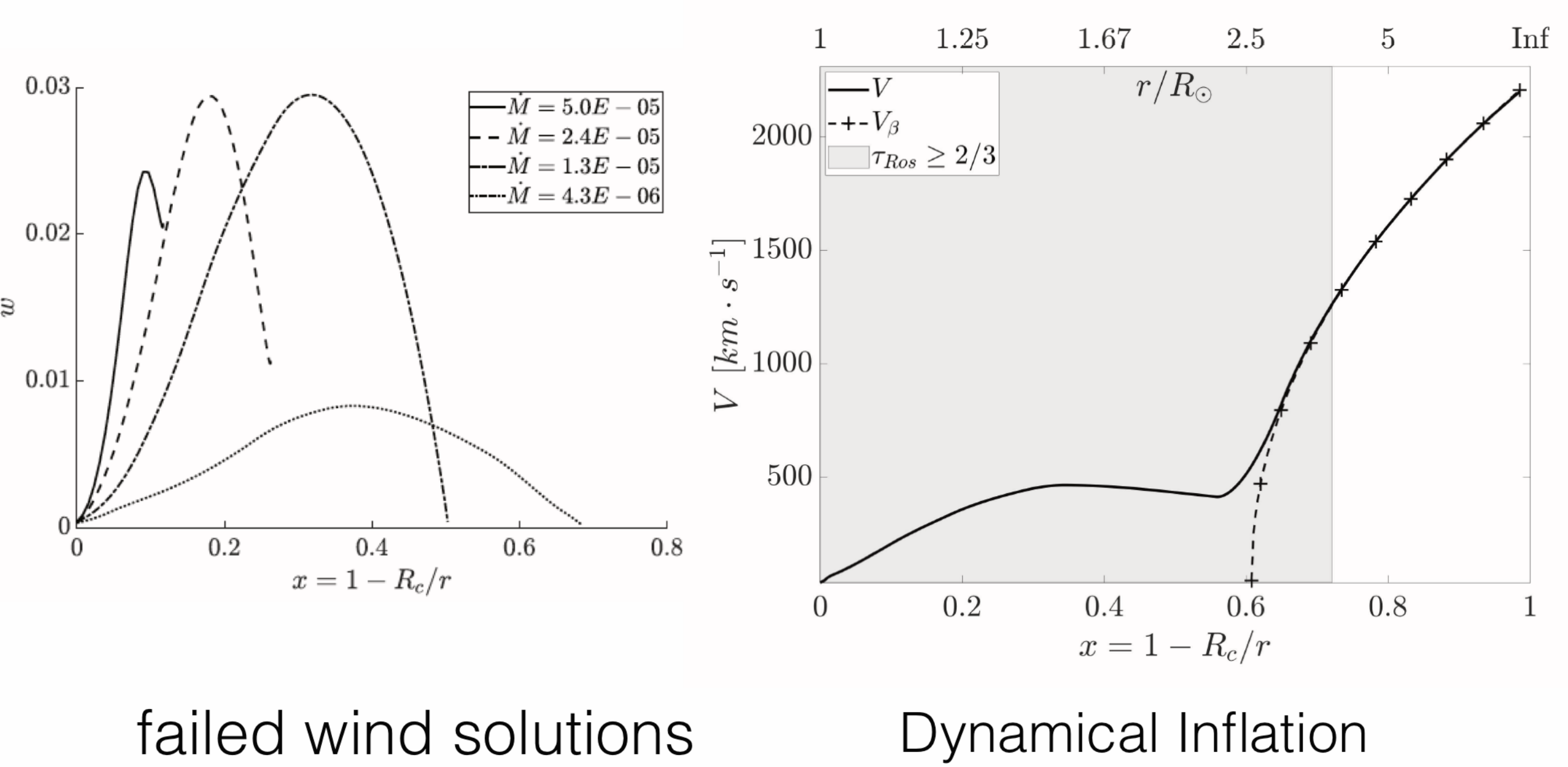} 
 \caption{
 Left: Gravitationally scaled flow energy $w=v^2/v_{esc}^2$ plotted vs. scaled radius $x=1-R_c/r$, where $R_c$ is the core radius near the wind sonic point.
 The various curves illustrate failed outflow solutions initiated at various densities by the iron opacity bump, with thus the various labeled mass
 loss rates, ${\dot M}$, in $M_\odot$/yr. 
 Right: Flow speed vs. scale radius $x$ for two-stage wind acceleration model. 
 Gray shading shows the optically thick wind region where iron-bump opacity initiates a wind outflow, which is sustained by including velocity desaturation that transitions to a CAK-like line-driven outflow, as fitted in the dashed plus-sign curve by a standard wind velocity law.
 This two-stage process can be characterized as a kind of ``dynamical inflation'' of the outer stellar envelope.
 Figures adopted from \cite[Poniatowski et al. (2021)]{Poniatowski21}.
 }
 \label{fig08}
\end{center}
\end{figure}

As illustrated in the left panel of figure \ref{fig08}, the decline in this iron bump opacity with decreasing temperature means that it alone cannot sustain such outflow driving in the cooler, outer layers,  leading then to a gravitational slowing and eventual stagnation of the wind outflow.
The right panel shows, however, that including the velocity desaturation of CAK-like line-driving can rekindle the outward acceleration and so sustain the outflow.

The overall two-stage model represents a kind of ``dynamical inflation'', with associated mass loss rates an order of magnitude higher than obtained in O-stars of comparable luminosity, in extreme cases approaching ${\dot M} \sim 10^{-4} M_\odot$/yr.
As illustrated in figure \ref{fig01}, it is somewhat akin to a water bubbler stream that is then taken up by the suction from a drinker.
Talks by N. Moens and A. Sander in these proceedings discuss further such WR wind acceleration.

\begin{figure}[b]
\begin{center}
\includegraphics[width=3.5in]{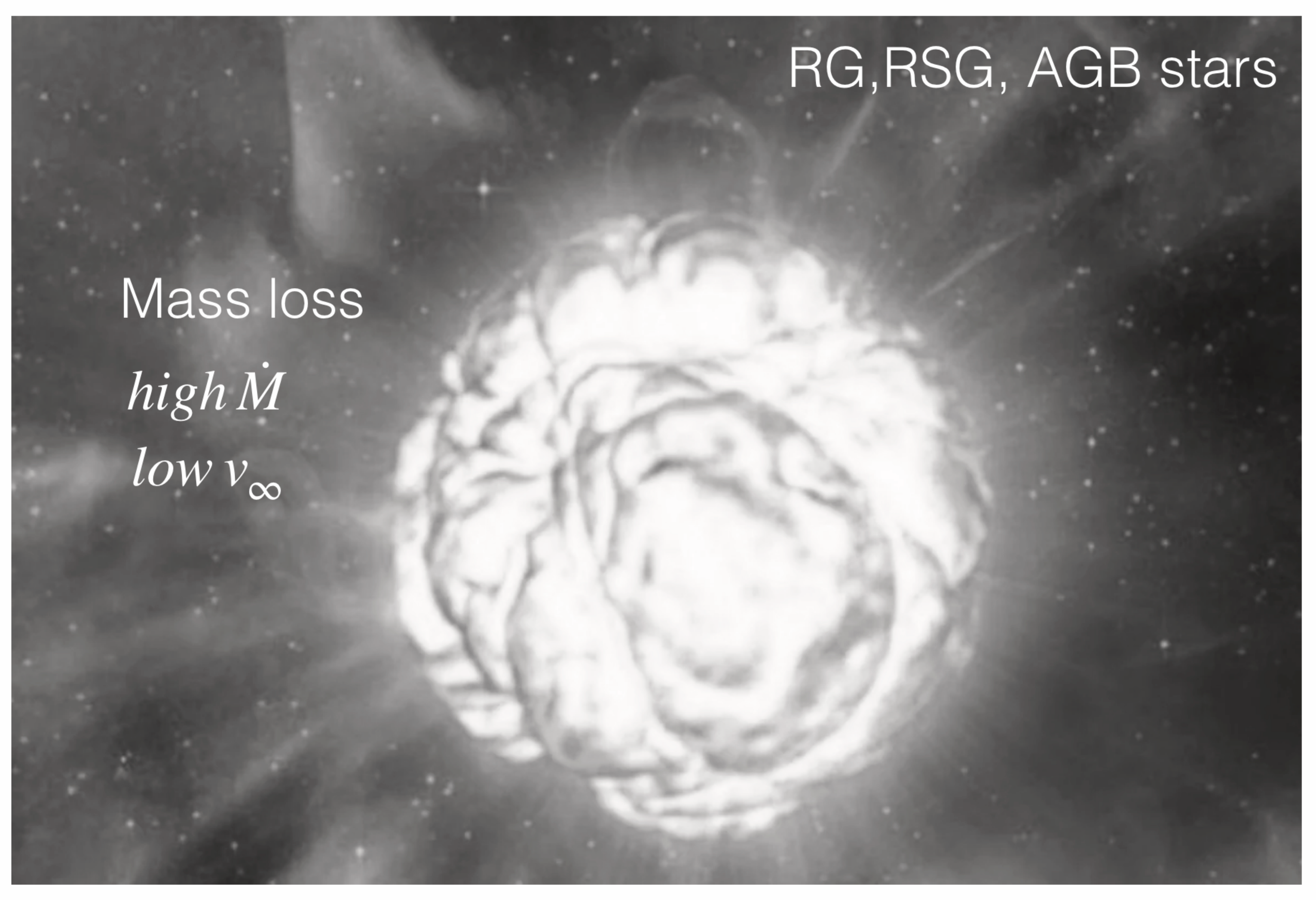} 
 \caption{Illustration of large convective plumes of a giant star, which can help initiate overflows mass loss with high mass loss rate ${\dot M}$ but low flow speeds $v_\infty$.}
   \label{fig09}
\end{center}
\end{figure}

\subsection{Slow overflow mass loss from giant stars}
Finally let us turn to the strong mass loss that can occur from evolved giant stars like Red (Super)Giants (RG and RSG) and those on the Asymptotic Giant Branch (AGB).
Like the overflowing glass illustrated in figure \ref{fig01}, this can be thought of a kind spillage, with overflowing material escaping the much weaker gravity of such stars, with flow speeds that barely exceed the reduced escape speed.
Even for the lower sound speed associated with a lower surface temperature $T_\ast \sim 3000$\,K, the larger radius now can lead to a moderate scale height to radius ratio
\beq
\frac{H}{R} = 0.2 \frac{T}{3000 {\rm K}} \, \frac{R/1000\,R_\odot}{M/M_\odot} 
\, ,
\label{eq:hbrgiants}
\eeq
Fgure \ref{fig08} shows how the associated convection cells in such stars can thus extend over a substantial fraction of a stellar radius.
As discussed in the contribution by S. Hoefner in these proceedings, detailed 3D simulations show this can sporadically suspend material to a large enough radius that the temperature becomes low enough ($T  \lesssim 1500$\,K) to initiate dust formation.
The high opacity of this dust can lead to a radiative acceleration that exceeds gravity, so driving material to full escape from the star.

For dust grains of radius $a$, density $\rho_d$ and mass fraction $X_d$, the dust opacity can be written as
\beq
\kappa_d = \frac{3 X_d}{4 a \rho_d} \approx 150 \frac{cm^2}{g} \, \frac{X_d}{X_{d\odot}} \, \frac{0.1 \mu m}{a}\,  \frac{1 \, g/cm^3}{\rho_d}
\, ,
\label{eq:kappad}
\eeq
where $X_{d\odot} = 0.002$ is the maximum dust fraction for solar abundance of dust-forming elements.

By comparison, the critical (Eddington) opacity to overcome gravity has the scaling (\cite[Hoefner and Olofsson 2018]{HO18}),
\beq
\kappa_{crit} = \frac{4 \pi GMc}{L} \approx  2.6 \frac{cm^2}{g} \, \frac{M}{M_\odot} \, \frac{5000 L_\odot}{L} 
\, .
\label{eq:kapc}
\eeq
Comparison shows that for luminous giant stars even partial dust formation should have an opacity that is sufficient to drive material to escape.

There remains some debate as to whether dust driving is essential to giant-star mass loss, or simply augments it.
For example, there are models (\cite[e.g., Kee et al. 2021]{Kee21}) for steady outflows driven by turbulent pressure, wherein  as in pressure-driven coronal models,
the mass loss rate is set by the density at the extended sonic point, but with values that greatly exceeds values of the solar wind,
as high as $10^{-6} M_\odot$/yr.
The presentation by J. Sundqvist in these proceedings provides further discussion of such turbulence-driven wind models.

\section{Eruptive mass loss}
\subsection{Energy requirement for eLBV mass ejection}
To complement the above discussion of the initiation of steady winds, let's next review the processes for initiating mass loss in an observational class of luminous, massive stars known as ``eruptive Luminous Blue Variables" (eLBV's).
Figure \ref{fig02} summarizes how these have properties that are intermediate between steady solar and stellar winds and the explosive mass ejection of core-collapse supernovae (SNe).
In the latter, the energy generated by core collapse deposits sufficient energy into the overlying stellar envelope to blow it completely from star, over an initial dynamical timescale of seconds, and with ejecta speeds that can approach $0.1$c.

Such SNe are thus at opposite extreme of the gradual, steady mass loss in winds, for which initiation occurs in surface layers when an outward force overcomes the inward gravitational acceleration. Even in the extreme cases of WR winds or mass loss from giant stars, the energy needed to escape is generally a small fraction of the available luminosity emitted by radiation.

The giant eruptiions seen from eLBV's have properties intermediate between winds and SNe.
While their initiation may be sudden, their evolution can extend over years or even decades, much longer than a dynamical timescale,
but generally short compared with a thermal relaxation timescale of the erupting star.
The mass fraction ejected can be up to 10-20\% of the stellar, much less than the full envelope ejection of SNe, but much larger than even the cumulative mass loss of stellar winds.
And unlike the fixed terminal speed of winds,  eLBV ejecta speed can have broad range, extending  from a faster, low density leading edge to a bulk of mass that just barely escapes the star's gravity.

A promising paradigm is to consider such eLBV eruptions as arising from a quite sudden addition energy to the star's envelope, which, unlike the explosive addition of SNe, is however only some fraction $f<1$ of the envelope binding energy.

\begin{figure}
\begin{center}
\includegraphics[width=5.5in]{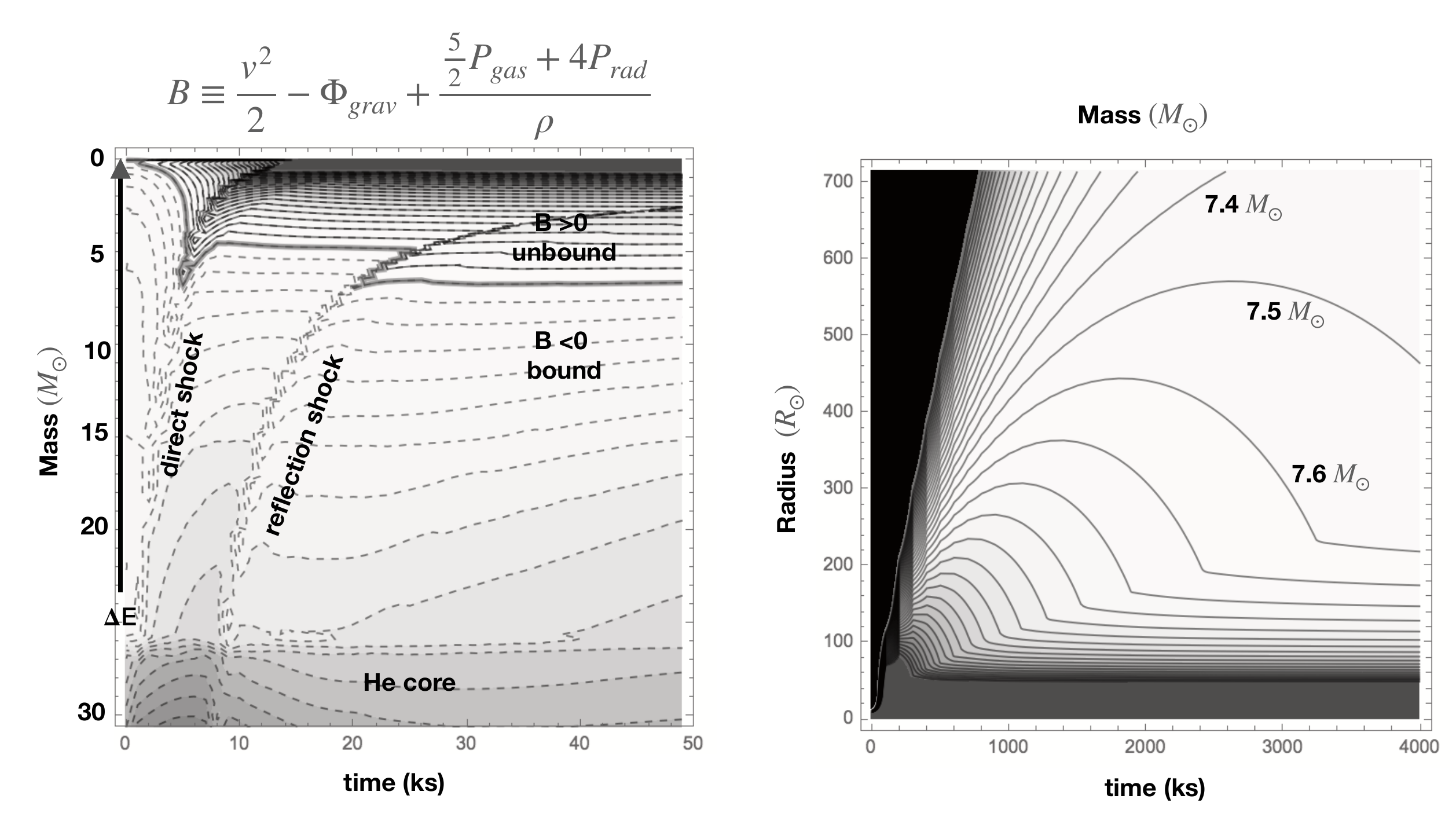} 
 \caption{
 Left:Contours of Bernouli energy $B$ as a function of  mass below surface and time since addition of an energy that is half the star's binding energy over the outer $25 M_\odot$. The resulting direct and reflected shocks heat the outer layers, making the outer $\sim 7 M_\odot$ become unbound with $B>0$.
 Right: The resulting outflow of the stellar envelope, with escape of $\sim 7.3 M_\odot$, with the rest of the material falling back onto the star.
 Figures adopted from \cite[Owocki et al. (2019)]{O19}.
 }
   \label{fig10}
\end{center}
\end{figure}

Figure \ref{fig10} show results for 1D hydrodynamical simulations of the response when a  energy that is a fraction $f=1/2	$ of the stellar
binding energy is addied to the  outer 25\,$M_\odot $ of a 100\,$M_\odot$ star, as indicated by the arrow along the left axis of the left panel.
In terms of fluid parcels defined by their mass coordinate  from the surface, the contours show the time response of the total ``Bernouli'' specific energy,
\beq
B \equiv  \frac{v^2}{2} - \Phi_{grav} + h
\, ,
\label{eq:Bdef}
\eeq
where $v$ is the flow speed, $\Phi_{grav}$ is the gravitational potential, and $h$ is the total specific enthalpy from gas and radiation;
 in terms of gas density and gas and radiation pressure, this is given by 
\beq
h = 
\frac{ \frac{5}{2} P_{gas} + 4 P_{rad}}{\rho}  = \frac{5}{2} \frac{kT}{\mu} + 
 \frac{ 4 a_{rad} T^4}{3 \rho}
\, ,
\label{eq:hdef}
\eeq
with $a_{rad}$ is the radiation constant.
Note how the initial energy addition induces a pair of direct and shock front that propagate toward the surface, heating the gas there so that the increased enthalpy makes the total Bernouli energy positive for the upper $\sim 7 M_\odot$ from the surface.
The result is an outward expansion of the stellar envelope, with the positive energy mass parcels with $M < 7.4 M_\odot$ escaping completely from the star, while the negative energy mass parcels with $M > 7.5 M_\odot$ eventually back onto the star.

 
The ejecta's variations in time $t$ and radius $r$ for the velocity $v$, density $\rho$, and temperature $T$ are quite well fit by similarity forms in the variable $r/t \approx v$. Specifically the scaled density follows a simple exponential decline $\rho t^3 \sim  \exp(-r/v_ot )$. 
This {\em exponential similarity} leads to analytic scaling relations for total ejecta mass $\Delta M$ and kinetic energy $\Delta K $ that agree well with the hydrodynamical simulations, with the specific-energy-averaged speed related to the exponential scale speed $v_o$ through 
${\bar v} = \sqrt{2 \Delta  K/ \Delta M }= \sqrt{12} v_o$, and a value comparable to the star's surface escape speed, $v_{esc}$. 

Unlike the fixed terminal speed $v_\infty$ of a stellar wind, a small amount of material can be ejected at very high speeds, $> 5000$\,km/s.
But like stellar winds, gravity still plays a central role in controlling the mass and speed of the outflow, through the ratio of the added energy to the gravitational binding energy.  This is distinct from standard SNe explosion, for which the added energy essentially overwhelms the gravitational binding of the envelope, leading to explosion speeds of order 0.1c

\begin{figure}[b]
\begin{center}
\includegraphics[width=3.5in]{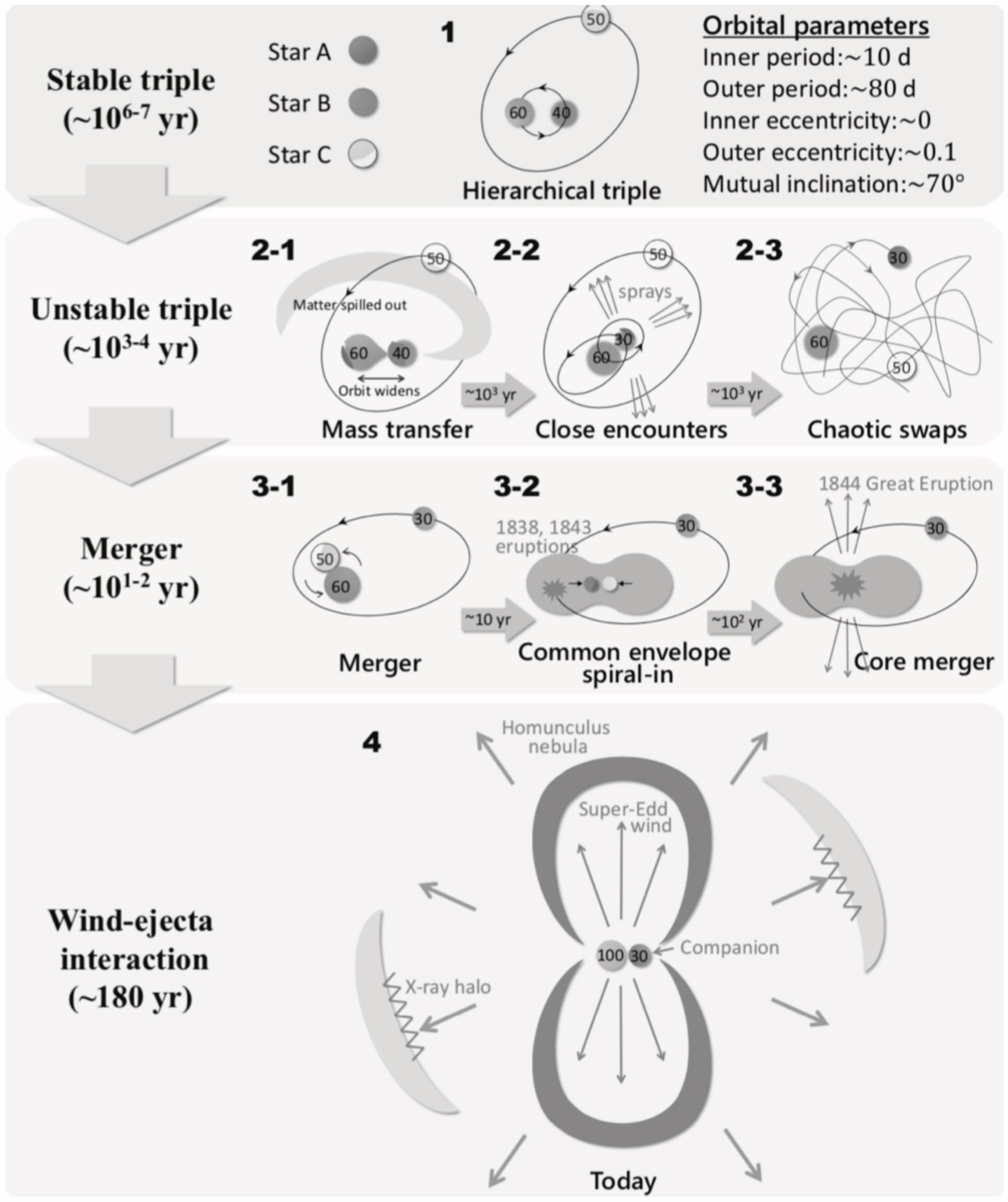} 
 \caption{Schematic outline of multi-step process leading to a binary merger that triggers the giant eruption of $\eta$~Carinae,
 with the post-eruption, bipolar , super-Eddington wind shaping the Homunculus nebula. Figure adopted from \cite[Hirai et al. (2021)]{Hirai21}.} 
   \label{fig11}
\end{center}
\end{figure}

\subsection{Merger model for $\eta$~Carinae}
The 1840's giant eruption of $\eta$~Carinae is perhaps the most prominent and extreme example of an eLBV, 
with the estimated $10-20 \, M_\odot$ ejected mass forming the bipolar `Homunculs' nebula (shown in the 3rd image in figure \ref{fig02}).
Two key challenges are to understand both the energy source powering the eruption, and the causes of the bipolar form.
$\eta$~Carinae is known to have a massive companion in an eccentric ($\epsilon \approx 0.9$ orbit with period $\sim 5.5$\,yr.

Figure \ref{fig11} summarizes a recently proposed model by \cite[Hirai et al. (2021)]{Hirai21}, in which an original triple system (phase 1) becomes unstable due to mass exchange (phase 2), leading to swaps and close encounters that result in a series of random ejecta that today are observed with
source times extending back several centuries.
Eventually, orbital decay of the innermost pair leads to a merger (phase 3), powering the 1840's giant eruption.
The enhanced luminosity and rapid rotation of the post-merger star (phase 4) drives a strong, bipolar, super-Eddington wind 
that sculpts the compressed Homonculus nebula seen today.

\section{Summary}
Within this symposium on ``The Origin of Outflows from Evolved Stars", a general overall theme for the above review 
on the initiation of such outflows is the key role played by gravity, and the need for forces that can overcome its inward pull to start an outflow, but also for this to be sustained to allow escape from the star's gravitational potential.

For coronal winds of the sun and other cool stars with convective envelopes, this is achieved by the high gas pressure associated with turbulent heating and thermal runway to temperatures, for which the gas internal energy becomes comparable to the gravitational binding.

For hot, massive OB stars, the momentum of the high stellar luminosity becomes effective tapped by the line-scattering for bound-bound transitions of heavy ions, with the resonantly enhanced opacity maintained by desaturation of the doppler shifted outflow;
this results in a sudden onset of the line-force near the sonic point, resulting in a line-driven suction of material from the star,
for which the details depend on the nature of scattering in this transonic region.

In the much stronger, optically thick winds of Wolf-Rayet stars, the wind blanketing increases the base temperature to be close enough to  that of the iron opacity bump, allowing a stronger wind initiation that is then sustained in the outer regions by the line-opacity desaturation.

In giant stars, the reduction in surface gravity and escape speed allow for giant convection cells to suspend material to levels with temperatures $T \lesssim 1500$\,K, cool enough to initiate dust formation; the coupling of this enhanced dust opacity to the high luminosity can then dirve material to full escape from the star. But even without dust formation, the turbulent pressure can drive significant mass outflows against the weak gravity.

For eLBV eruptions, a key again is to provide an energy deposition to allow some fraction of the stellar envelope to gain sufficient energy to be ejected from the star.  Instead of a steady wind, each mass parcel follows a trajectory tied to its total energy, with now a distribution of escape speeds that are described by a similarity form, with fall back to the star for bulk of material with net negative total energy.

While the outline here presents an idealized view of steady vs. impulsive mass loss, in practice real outflows will exhibit combinations of such traits, e.g. with infall in some WR winds, and with eLBV eruptions punctuated by quasi-steady super-Eddington winds.
The further contributions in these proceeding explore in greater detail the many variations on these themes that occur in the outflows from evolved stars.

\section*{Acknowledgments}
I thank the IAUS366 SOC, and particularly its chair Leen Decin, for the invitation to give this opening review.
I also thank L. Poniatowski and R. Hirai for permission to use figures \ref{fig08} and \ref{fig11}, respectively.


\begin{discussion}

\discuss{Decin}{In your discussion of eruptive mass loss, what fraction of the stellar mass can be lost in such eruptions?}

\discuss{Owocki}
{It depends on both the location of the energy addition and its fraction of the stellar binding energy.  The specific model I discussed ejected about 7\% of the mass of the  star's 100$M_\odot$, but a parameter study shows it is possible to eject more or less than this. A large energy added in the stellar core can eject the entire stellar envelope, as in a SNe explosion. Energy addition to a near-surface layer will at most eject the mass in that layer.
Energy from a merger seems sufficient to eject $>10$\%  of a star's mass, as inferred for $\eta$~Carinae.}

\discuss{Mellah}{I had a question about the bipolar outflow, with the polar outflow being faster and more dilute}

\discuss{Owocki}{No, it's not more dilute, because in a rapidly rotating radiative envelope, the bright poles can drive both a faster and denser outflow.
In this model, that's how the Homunculus shape is formed.}

\discuss{Mellah}{So you wouid not see the material in the orbital plane.}

\discuss{Owocki}{Well, there is some material ejected mechanically in the rotational/orbital plane, and this may be the origin of the observed equatorial skirt.}

\discuss{De Marco}{In a binary, there's of course the straight gravity of the companion to help a star to lose mass. 
But do you envision a way in which a compansion can trigger mass loss in a way that a star uses its own reservoir of energy?}

\discuss{Owocki}
{Well there are some ideas, including I believe by yourself, for how orbital motion can excite pulsation, with perhaps certain resonances, resulting then
in episodic mass ejection.
But to get a {\em giant} eruption, where, as in $\eta$~Carinae, the star loses 10\% or more of its mass, I think you need to add a large source of 
energy, either externally from a merger, or some kind of ignition of enhanced burning in the core.}

\discuss{Sahai}{In regards to mass loss in binaries, I wanted to point out there's a beautiful example in our Galaxy called the Boomerang nebula, wherein there's a huge amount of mass ejected from an intermediate mass star with relatively high speeds.
An other example is V Hydra, where one sees bullet-like ejection triggered from periastron passage of a companion.}

\discuss{Owocki}{Yes, I didn't intend my talk to be comprehensive, and indeed the main focus was on initiation of wind mass loss. But certainly 
the topic of eruptive mass loss, and the role of binarity, is a broad area worthy of much further discussion in this symposium.}

\end{discussion}


\begin{thebibliography}{}

\bibitem[Castor et al.(1975)]{CAK75} Castor, J.~I., Abbott, D.~C., \& Klein, R.~I.\ 1975, \apj.
\bibitem[Gayley(1995)]{Gayley95} Gayley, K.~G.\ 1995, \apj, 454, 410.
\bibitem[Gr{\"a}fener et al.(2012)]{Graefener12} Gr{\"a}fener, G., Owocki, S.~P., \& Vink, J.~S.\ 2012, \aap, 538, A40. 
\bibitem[Hirai et al.(2021)]{Hirai21} Hirai, R., Podsiadlowski, P., Owocki, S.~P., et al.\ 2021, \mnras, 503, 4276. 
\bibitem[H{\"o}fner \& Olofsson(2018)]{HO18} H{\"o}fner, S. \& Olofsson, H.\ 2018, \aapr, 26, 1. 
\bibitem[Kee et al.(2021)]{Kee21 } Kee, N.~D., Sundqvist, J.~O., Decin, L., et al.\ 2021, \aap, 646, A180. 
\bibitem[Owocki \& Puls(1996)]{OP96} Owocki, S.~P. \& Puls, J.\ 1996, \apj. 462, 894. 
\bibitem[Owocki \& Puls(1999)]{OP99} Owocki, S.~P. \& Puls, J.\ 1999, \apj, 510, 355. 
\bibitem[Owocki et al.(2019)]{O19} Owocki, S.~P., Hirai, R., Podsiadlowski, P., et al.\ 2019, \mnras, 485, 988. 
\bibitem[Poniatowski et al.(2021)]{Poniatowski21} Poniatowski, L.~G., Sundqvist, J.~O., Kee, N.~D., et al.\ 2021, \aap, 647, A151. 
\bibitem[Sobolev(1960)]{Sobolev60} Sobolev, V.~V.\ 1960, Cambridge: Harvard University Press, 1960.

\end{thebibliography}
\end{document}